\documentclass[floats,floatfix,showpacs,amssymb,prd,twocolumn,superscriptaddress,nofootinbib]{revtex4-1}

\usepackage{amssymb,amsmath,verbatim,mathtools,needspace,enumitem,etoolbox,graphicx,physics,microtype,afterpage,bm}

\usepackage{subcaption}
\usepackage{ragged2e}
\DeclareCaptionJustification{justified}{\justifying}
\usepackage[dvipsnames, usenames]{xcolor}
\usepackage[normalem]{ulem}

\definecolor{linkcolor}{rgb}{0.0,0.3,0.5}
\usepackage[unicode, colorlinks=true, linkcolor=linkcolor, citecolor=linkcolor, filecolor=linkcolor,urlcolor=linkcolor, pdfusetitle]{hyperref}
\usepackage[all]{hypcap}
\usepackage[T1]{fontenc}
\usepackage[utf8]{inputenc}
\usepackage{tabularx}
\usepackage{float}
\allowdisplaybreaks
\renewcommand{\arraystretch}{1.4}

\graphicspath{{./figure/}}

\newcommand{\llp}{\left [}
\newcommand{\rrp}{\right ]}

\newcommand{\co}{\text{\tiny cut-off}}

\def\PBH{\text{\tiny  PBH}}

\newcommand{\be}{\begin{equation}\begin{aligned}}
\newcommand{\ee}{\end{aligned}\end{equation}}

\newcommand{\bbe}{\begin{align}}
\newcommand{\eee}{\end{align}}

\newcommand{\bea}{\begin{eqnarray}}
\newcommand{\eea}{\end{eqnarray}}

\def\beq{\begin{equation}}
\def\eeq{\end{equation}}

\def\d{{\rm d}}

\def\beqa{\begin{eqnarray}}

	\def\eeqa{\end{eqnarray}}
	
\def\lsim{\mathrel{\rlap{\lower4pt\hbox{\hskip0.5pt$\sim$}}
		\raise1pt\hbox{$<$}}}     
\def\gsim{\mathrel{\rlap{\lower4pt\hbox{\hskip0.5pt$\sim$}}
		\raise1pt\hbox{$>$}}}     

\def\d{{\rm d}}

\def\d{{\rm d}}

\def\PBH{\text{\tiny \rm PBH}}

\newcommand{\DM}{\text{\tiny DM}}
\def\eeqa{\end{eqnarray}}

\def\bq{\begin{quote}}
\def\eq{\end{quote}}

 at 10truept

\def\eeqa{\end{eqnarray}}
\def\lsim{\mathrel{\rlap{\lower4pt\hbox{\hskip0.5pt$\sim$}}
  \raise1pt\hbox{$<$}}}     
\def\gsim{\mathrel{\rlap{\lower4pt\hbox{\hskip0.5pt$\sim$}}
  \raise1pt\hbox{$>$}}}     

\def\msun{\ M_\odot}

\newcommand{\spp}{ \hspace{.0 cm}}

\def\apjs{\rm{ApJS}}

\def\pasp{\rm{PASP}}

\newcommand{\jhu}{\affiliation{Department of Physics and Astronomy, Johns Hopkins University, 3400 N. Charles
Street, Baltimore, MD 21218, USA}}

\definecolor{rb4}{HTML}{27408B}

\begin{document}

\author{Gabriele~Franciolini}
\email{Gabriele.Franciolini@uniroma1.it}
\affiliation{D\'epartement de Physique Th\'eorique and Centre for Astroparticle Physics (CAP), Universit\'e de Gen\`eve, 24 quai E. Ansermet, CH-1211 Geneva, Switzerland}
\affiliation{Dipartimento di Fisica, Sapienza Università 
di Roma, Piazzale Aldo Moro 5, 00185, Roma, Italy}

\author{Vishal~Baibhav} 
\jhu

\author{Valerio~De~Luca}
\affiliation{D\'epartement de Physique Th\'eorique and Centre for Astroparticle Physics (CAP), Universit\'e de Gen\`eve, 24 quai E. Ansermet, CH-1211 Geneva, Switzerland}
\affiliation{Dipartimento di Fisica, Sapienza Università 
di Roma, Piazzale Aldo Moro 5, 00185, Roma, Italy}

\author{Ken~K.~Y.~Ng}
\affiliation{LIGO Laboratory, Massachusetts Institute of Technology, Cambridge, Massachusetts 02139, USA}
\affiliation{Kavli Institute for Astrophysics and Space Research, Massachusetts Institute of Technology, Cambridge, Massachusetts 02139, USA}

\author{Kaze~W.~K.~Wong} 
\jhu

\author{Emanuele~Berti} 
\jhu

\author{Paolo~Pani}
\affiliation{Dipartimento di Fisica, Sapienza Università 
di Roma, Piazzale Aldo Moro 5, 00185, Roma, Italy}
\affiliation{INFN, Sezione di Roma, Piazzale Aldo Moro 2, 00185, Roma, Italy,}

\author{Antonio~Riotto}
\affiliation{D\'epartement de Physique Th\'eorique and Centre for Astroparticle Physics (CAP), Universit\'e de Gen\`eve, 24 quai E. Ansermet, CH-1211 Geneva, Switzerland}

\author{Salvatore~Vitale}
\affiliation{LIGO Laboratory, Massachusetts Institute of Technology, Cambridge, Massachusetts 02139, USA}
\affiliation{Kavli Institute for Astrophysics and Space Research, Massachusetts Institute of Technology, Cambridge, Massachusetts 02139, USA}


\title{Searching for a subpopulation of primordial black holes \\ in LIGO/Virgo gravitational-wave data}

\begin{abstract} \noindent

With several dozen binary black hole events detected by LIGO/Virgo to date and many more expected in the next few years, gravitational-wave astronomy is shifting from individual-event analyses to population studies. Using the GWTC-2 catalog, we perform a hierarchical Bayesian analysis that for the first time combines several state-of-the-art astrophysical formation models with a population of primordial black holes~(PBHs)
and constrains the fraction of a putative subpopulation of PBHs in the data. 
We find that this fraction depends significantly on the set of assumed astrophysical models.
While a primordial population is statistically favored against certain competitive astrophysical channels, such as globular clusters and nuclear stellar clusters, a dominant contribution from the stable-mass-transfer isolated formation channel drastically reduces the need for PBHs, except for explaining the rate of mass-gap events like GW190521. 
The tantalizing possibility that black holes formed after inflation are contributing to LIGO/Virgo observations  could only be verified by further reducing uncertainties in astrophysical and primordial formation models, and it may ultimately be confirmed by third-generation interferometers.

\end{abstract}

\maketitle

\noindent{{\bf{\em Introduction.}}}
The second gravitational-wave transient catalog (GWTC-2)~\cite{Abbott:2020niy,Abbott:2020gyp} of compact binary mergers detected by the LIGO/Virgo collaboration~(LVC)~\cite{TheLIGOScientific:2014jea,TheVirgo:2014hva} brought the total number of binary black holes~(BBHs) reported by the LVC  to 47.
Recently, this figure increased further with the latest GWTC catalog~\cite{LIGOScientific:2021djp}.
Additional detections have been reported by independent groups using public data, though usually with lower statistical significance (see e.g.~\cite{Nitz:2019hdf,Zackay:2019btq,Roulet:2020wyq}). 
As the number of observations increases, we can characterize with increasing accuracy the properties of the underlying population of black holes~(BHs) and the relative contribution of various BBH formation channels.

In their population analysis, the LVC has used phenomenological models built to capture key expected features of the mass, spin, and redshift distribution of BBHs (e.g. a power-law mass distribution), but not the physical mechanisms responsible for these features (e.g., mass transfer in binary evolution)~\cite{Abbott:2020gyp,LIGOScientific:2021psn}.
The model that is preferred by GWTC-2 data describes the distribution of the primary (i.e., most massive) BH in the binary as the sum of a power-law and a Gaussian distribution, denoted as ``Power Law + Peak'' in Ref.~\cite{Abbott:2020gyp}. The model has several free parameters
and it is preferred to a simpler power-law function, which might suggest that multiple formation channels are at play.

Many astrophysical formation scenarios could contribute to the observed population~\cite{Mandel:2018hfr,Mapelli:2018uds}. The observed excess of massive BHs could be the result of hierarchical mergers of smaller objects~\cite{Gerosa:2017kvu,Fishbach:2017dwv,Rodriguez:2019huv,Gerosa:2019zmo,Gerosa:2021hsc}, the end product of the life of massive stars just below the pair-instability supernova mass gap~\cite{Heger:2001cd,Belczynski:2016jno,Gayathri:2021xwb}, or it may be of primordial origin~\cite{DeLuca:2020sae,Kritos:2020wcl}.
One event in particular, GW190521~\cite{Abbott:2020tfl}, challenges traditional formation scenarios. With component masses of $m_1=90.9^{+29.1}_{-17.3}\msun{}$ and $m_2=66.3^{+19.3}_{-20.3}\msun{}$, GW190521 is the most massive BBH detected to date. The posterior of the primary mass has support nearly entirely in the pair-instability supernova mass gap, where BHs are not expected to form from the collapse of massive stars (see~\cite{Farmer:2019jed,Sakstein:2020axg,Belczynski:2020bca,Costa:2020xbc,Tanikawa:2020abs,Woosley:2021xba,Baxter:2021swn} for discussions of astrophysical uncertainties in this prediction). 

In addition to astrophysical formation channels, a tantalizing possibility is that a fraction of these events may be due to primordial BHs~(PBHs)~\cite{Zeldovich:1967lct,Hawking:1974rv,Chapline:1975ojl,Carr:1975qj} formed from the collapse of large overdensities in the radiation-dominated early universe~\cite{Ivanov:1994pa,GarciaBellido:1996qt,Ivanov:1997ia,Blinnikov:2016bxu}. In this scenario, PBHs are not clustered at formation \cite{Ali-Haimoud:2018dau,Desjacques:2018wuu,Ballesteros:2018swv,MoradinezhadDizgah:2019wjf,Inman:2019wvr,DeLuca:2020jug} and  primordial BBHs are assembled via gravitational decoupling from the Hubble flow before the matter-radiation equality \cite{Nakamura:1997sm,Ioka:1998nz} (see~\cite{Sasaki:2018dmp,Green:2020jor} for reviews). 
PBHs in different mass ranges could contribute to a sizeable fraction $f_\PBH\equiv \Omega_\PBH/\Omega_\DM$ of the dark matter energy density~\cite{Carr:2020gox}, but current GW data imply an upper bound $f_\PBH\lesssim {\cal O}(10^{-3})$ in the mass range of interest to current GW detectors~\cite{Bird:2016dcv,Clesse:2016vqa,Sasaki:2016jop,Eroshenko:2016hmn, Wang:2016ana,
Ali-Haimoud:2017rtz,Clesse:2017bsw, Chen:2018czv,Raidal:2018bbj, Hutsi:2019hlw, Vaskonen:2019jpv, Gow:2019pok,Wu:2020drm,DeLuca:2020bjf, Hall:2020daa,Wong:2020yig,Hutsi:2020sol,DeLuca:2021wjr,Deng:2021ezy,Kimura:2021sqz}.
A different scenario predicts that PBHs may form with a broad mass distribution shaped by the QCD transition~\cite{Jedamzik:1996mr,Byrnes:2018clq}, and could assemble dynamically in dense halos in the late-time universe~\cite{Carr:2019kxo,Jedamzik:2020omx,Clesse:2020ghq}. This, however, requires PBHs to be strongly clustered to evade existing astrophysical constraints on their abundance~\cite{Carr:2020gox}.

Overall, the data indicate that not all BBH events detected so far can be explained by a single formation channel, be it either astrophysical~\cite{Zevin:2020gbd} or primordial~\cite{Hall:2020daa} (see~\cite{Wong:2020yig} for the most updated analysis of the PBH scenario).  
Previous work tried to infer the mixing fraction of multiple astrophysical populations~\cite{Zevin:2017evb,Bouffanais:2019nrw,Wong:2020ise,Kimball:2020qyd,Zevin:2020gbd} and compared the PBH scenario against the phenomenological LVC power-law model~\cite{Hall:2020daa,Hutsi:2020sol,DeLuca:2021wjr}.
In this paper we present a more comprehensive hierarchical Bayesian inference study of the GWTC-2 catalog. For the first time, we mix the PBH model~\cite{Raidal:2018bbj,DeLuca:2020qqa} 
with several state-of-the-art astrophysical models that can reproduce many features of the observed population. 
This allows us to place statistical constraints on a putative subpopulation of PBHs in GW data given our present (admittedly incomplete) knowledge of BBH formation scenarios.

\noindent{{\bf{\em BBH models.}}}
Our astrophysical models come from Ref.~\cite{Zevin:2020gbd}, the most comprehensive attempt to date at comparing different astrophysical formation scenarios against LVC data. That work considered three field formation models and two dynamical formation models.
Among the three field formation scenarios -- a late-phase common envelope (CE), binaries that only have stable mass transfer between the
star and the already formed BH (SMT), and chemically
homogeneous evolution (CHE) -- Ref.~\cite{Zevin:2020gbd} found that the dominant channels correspond to the CE and SMT scenarios.
These two channels were simulated using the POSYDON framework~\cite{Bavera:2020uch,frago}, which models binary evolution with the population synthesis code COSMIC~\cite{Breivik:2019lmt} and uses MESA~\cite{2019ApJS..243...10P} for binary evolution calculations.  
The key parameters of these models are the CE efficiency $\alpha_\text{\tiny CE} \in [0.2, 0.5, 1.0, 2.0, 5.0]$, with large values of $\alpha_\text{\tiny CE}$
leading to efficient CE evolution, and the natal BH spin $\chi_{\rm b} \in [0, 0.1, 0.2, 0.5]$.
The two dynamical models consider formation in old, metal-poor globular clusters~(GC) and in nuclear star clusters~(NSC). The GC models are taken from a grid of 96 $N$-body models of collisional star clusters simulated using the cluster Monte Carlo code CMC~\cite{Rodriguez:2019huv}. The 96 models consist of four independent grids of 24 models, each with different initial spins. Large natal spins imply larger ejection probabilities, and therefore a smaller probability of repeated mergers. The NSC models
use COSMIC to generate the BH masses from a single stellar population with metallicity $Z=(0.01,0.1,1)Z_\odot$, and evolve the clusters and their BHs using the semianalytical approach described in~\cite{Antonini:2018auk}. This procedure is repeated for all values of $\chi_{\rm b}$ listed above.

For the PBH model, we compute merger rates following Refs.~\cite{Raidal:2018bbj, Vaskonen:2019jpv,DeLuca:2020jug,DeLuca:2020qqa}, as in other recent studies~\cite{DeLuca:2020sae,Wong:2020yig,DeLuca:2021wjr}.
The (lognormal) PBH mass function is characterized by a central mass $M_c$ (not to be confused with a binary's chirp mass ${\cal M}$) and a width $\sigma$.
Another hyperparameter is the PBH abundance $f_\PBH$.
Finally, PBHs may experience a phase of matter accretion during their cosmic evolution, impacting their mass and spin distributions at detection. 
As PBHs form from the collapse of radiation density perturbations in the early universe~\cite{Mirbabayi:2019uph, DeLuca:2019buf},
their natal spins are negligible and independent of $\chi_{\rm b}$. 
To capture uncertainties in the accretion model we introduce a cut-off redshift $z_\co$ below which accretion is inefficient. If $z_\co \gtrsim 30$, accretion is negligible in the mass range of interest for LVC observations and PBHs retain small spins even at low redshift, whereas $z_\co \simeq 10$ would correspond to a strong accretion phase, leading to larger PBH masses and spins~\cite{DeLuca:2020bjf,DeLuca:2020qqa}.
Similarly to the dynamical astrophysical channels, the PBH spin orientations with respect to the binary's angular momentum are expected to be independent and uniformly distributed on the sphere. 

Overall, our astrophysical models depend on the \emph{hyperparameters} 
${\bm \lambda}_\text{\tiny ABH}
=[\alpha_\text{\tiny CE},\chi_{\rm b},
N_\text{\tiny CE},N_\text{\tiny SMT},
N_\text{\tiny GC},
N_\text{\tiny NSC}]$, 
where the number of events in each channel $N_i$, following Ref.~\cite{Zevin:2020gbd}, is assumed to be unconstrained and independent of $\alpha_\text{\tiny CE}$ and $\chi_{\rm b}$.
The PBH channel depends on ${\bm \lambda}_\PBH=[M_c,\,\sigma,\,f_{\PBH},\,z_\co]$, with $N_\text{\tiny PBH}\approx f_\text{\tiny PBH}^2$~\cite{Wong:2020yig}. 

\begin{figure*}[th] 
\centering
\includegraphics[width=0.385\textwidth]{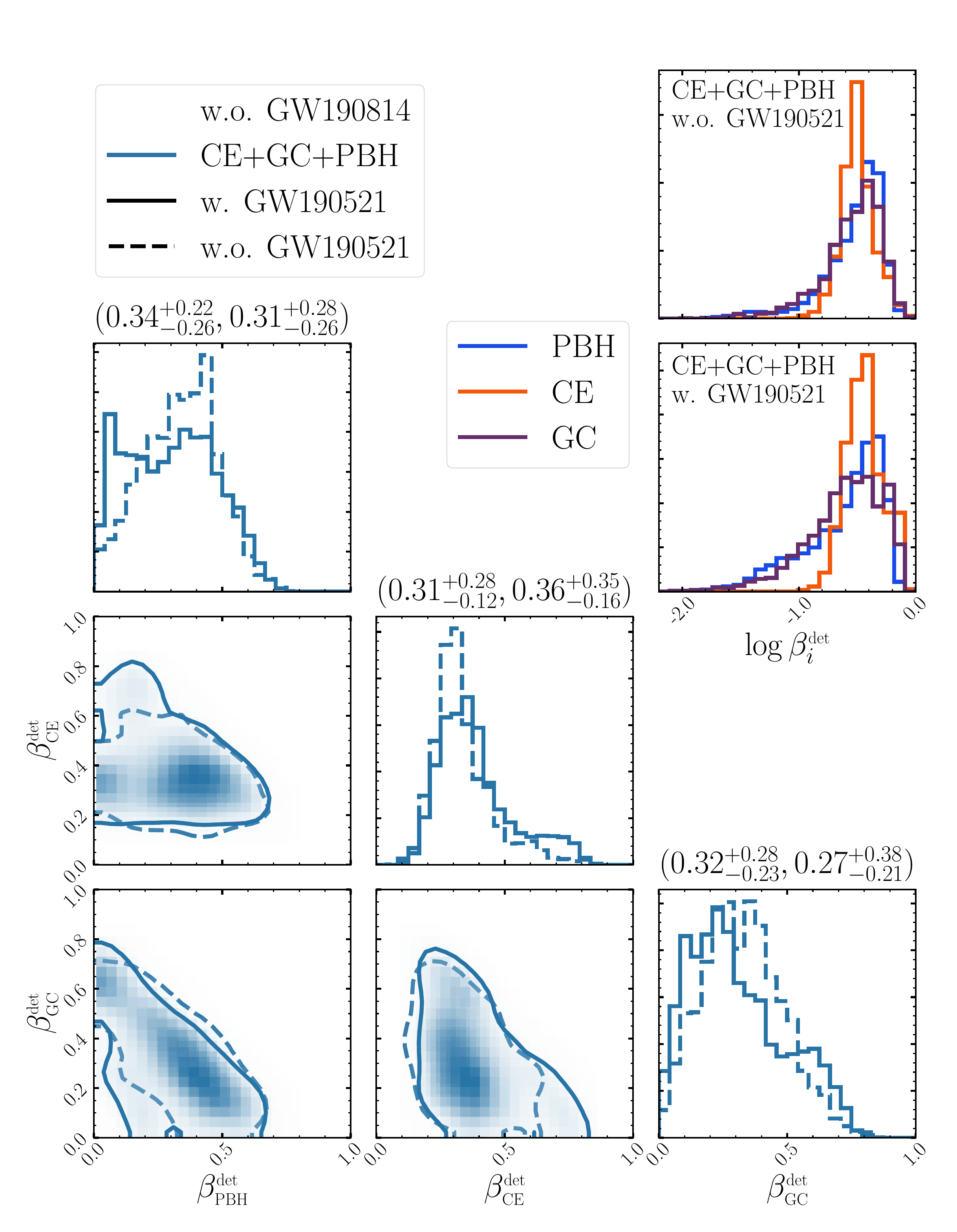}
\includegraphics[width=0.605\textwidth]{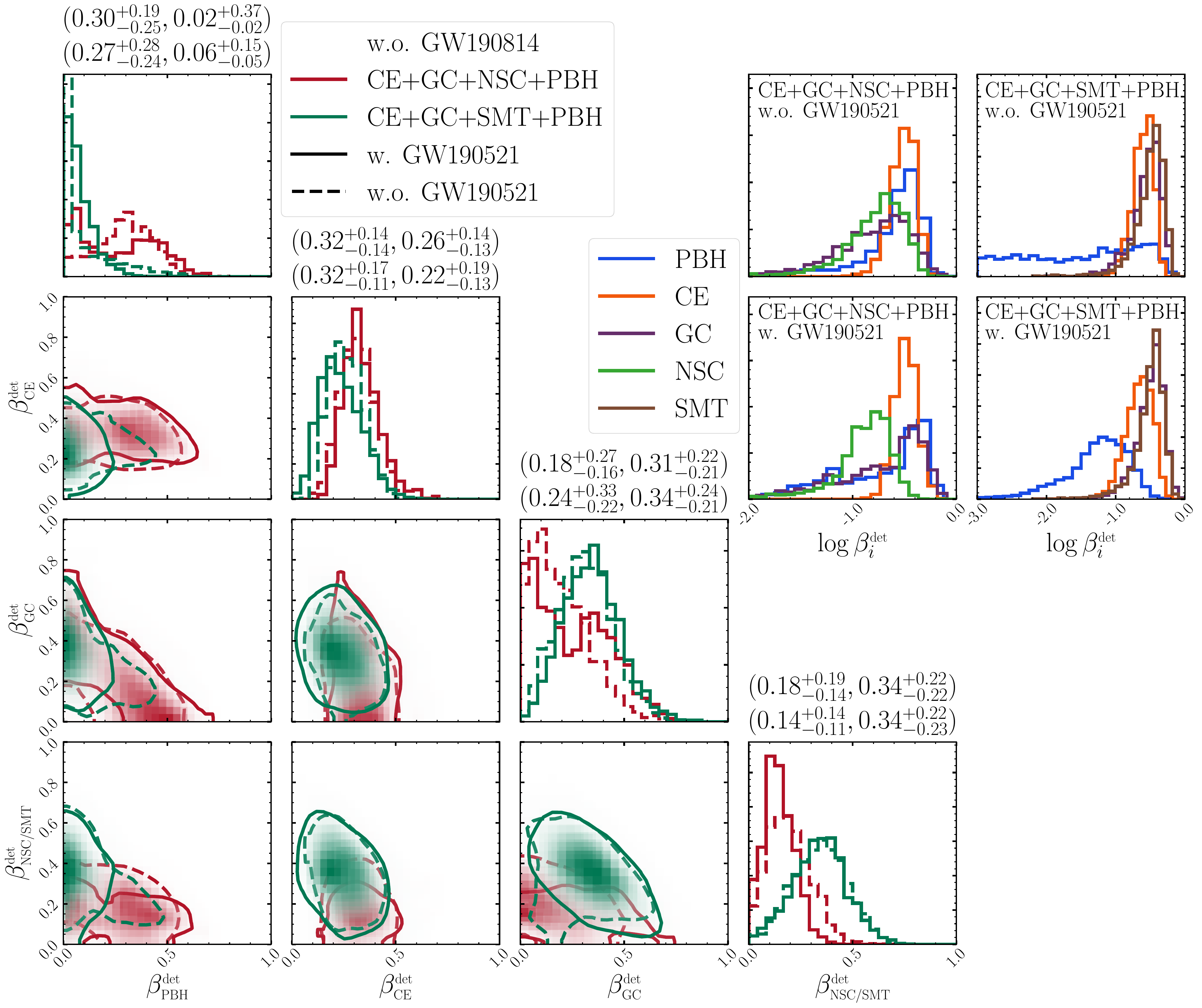}
  \caption{
  Posterior distributions of the individual {\it detectable} mixing fractions $\beta_{i}^\text{\tiny det}$ of different populations. The left panel refers to a 2+1 (CE+GC+PBH) model, whereas the right panel refers to two 3+1 models (CE+GC+NSC+PBH and CE+GC+SMT+PBH). In all cases we excluded GW190814, and we present results with and without including GW190521.
 The insets in the top right of each corner show each $\beta_{i}^\text{\tiny det}$ in logarithmic scale to highlight the monotonic behavior as $\beta_{i}^\text{\tiny det}\to 0$. The two $90\%$ confidence intervals (C.I.) for $\beta_i^\text{\tiny det}$ reported above each column in the left panel correspond to the models in the inset (from top to bottom); those on the right panel are sorted in the same way as the four panels in the inset. The corresponding posteriors for the  PBH hyperparameters are shown in the SM.
 }
\label{fig:corner}
\end{figure*}

\noindent{{\bf{\em Data analysis.}}}
Our setup follows Refs.~\cite{Wong:2020yig,DeLuca:2021wjr} and the inference is performed by sampling the likelihood~\cite{Vitale:2020aaz}
\begin{align}
p(\bm{\lambda}|\bm{d}) =
 \pi(\bm{\lambda})e^{- N_\text{\tiny det} ({\bm \lambda})} 
 \llp N({\bm \lambda})\rrp^{N_{\rm obs}}
 \prod_{i=1}^{N_{\rm obs}}
  \frac{1}{{\cal S}_i}
 \sum_{j=1}^{{\cal S}_i} \frac{p_\text{\tiny pop}(^j\bm{\theta}_i|\bm{\lambda})}{\pi(^j\bm{\theta}_i)}
\label{eq:populationPosterior_discrete}
\end{align}
in the space of 
${\bm \lambda}={\bm \lambda}_\text{\tiny ABH} \cup {\bm \lambda}_\text{\tiny PBH}$ by using the Markov chain Monte Carlo software \texttt{emcee}~\cite{2013PASP..125..306F}.
In Eq.~\eqref{eq:populationPosterior_discrete},
$N_\text{\tiny obs}$ is the number of GW events in the catalog;
$N({\bm \lambda})$ is the number of events in the model;
$N_\text{\tiny det} ({\bm \lambda})$ is the number of {\it observable} events computed by accounting for the experimental selection bias;
 ${\cal S}_i$ is the length of the posterior sample of each event in the catalog;
$\pi(\bm{\theta})$ is the prior on the binary parameters $\bm{\theta}$ used by the LVC when performing the parameter estimation --~this prior is removed to extract the values of the single-event likelihood, ensuring only the informative part of the event posterior is used and does not affect the population inference (but see Refs.~\cite{Pankow:2016udj, Vitale:2017cfs, Zevin:2020gxf, Bhagwat:2020bzh} for its impact on the interpretation of single events);
and $\pi(\bm{\lambda})$ is the prior on the hyperparameters, which is assumed to be flat.

The quantity $p_\text{\tiny pop}(\bm{\theta}|\bm{\lambda})$ is the distribution of the BBH parameters $\theta_i =[ m_1,m_2,z,\chi_\text{\tiny eff}]$, where $m_i$ is the source-frame  mass of the $i$-th binary component, $z$ is the merger redshift, and $\chi_\text{\tiny eff} \equiv (\chi_1 \cos{\alpha_1} + q \chi_2 \cos{\alpha_2})/(1+q)$ is the effective spin parameter, which is a function of the mass ratio $q\equiv m_2/m_1\leq 1$, of both BH spin magnitudes $\chi_j$ ($j=1,2$, with $0\leq \chi_j\leq 1$), and of their orientation with respect to the orbital angular momentum, parametrized by the tilt angles $\alpha_j$. In the inference we neglect the precessional spin $\chi_\text{\tiny p}$, since this parameter is poorly determined for most of the GW events detected to date~\cite{Abbott:2020gyp}. The Appendix gives more details on the calculation of $p_\text{\tiny pop}(\bm{\theta}|\bm{\lambda})$.

To statistically quantify the fraction of various channels given the GWTC-2 dataset, we compute the Bayes factor between model ${\cal M}_1$ and model ${\cal M}_2$, namely ${\cal B}^{{\cal M}_1}_{{\cal M}_2} \equiv { Z_{{\cal M}_1}}/{Z_{{\cal M}_2}}$, where $Z_{\cal M} \equiv \int \d {\bm \lambda} \, p(\bm{\lambda}|\bm{d})$
is the evidence. According to Jeffreys' scale criterion~\cite{Jeffreys}, a Bayes factor larger than $(10,10^{1.5},10^2)$ would imply a strong, very strong, or decisive Bayesian evidence in favor of model ${\cal M}_1$ with respect to model ${\cal M}_2$, given the available data.

\begin{table*}[th!]
\renewcommand{\arraystretch}{1.3}
\caption{
Bayesian evidence ratios for the different mixed astrophysical and primordial populations (normalised with respect to the CE+GC scenario), obtained by marginalising over $\alpha_\text{\tiny \rm CE}$ and $\chi_\text{\tiny \rm b}$,  with and without GW190521.}
\begin{tabularx}{2. \columnwidth}{Xccccc}
\hline
  \hline
$\log_{10} {\cal B}^{\cal  M}_{ \text{\tiny CE}+\text{\tiny GC}} $   \spp & 
 \spp  { CE+GC+PBH }\spp &
     \spp  { CE+GC+NSC} \spp   & 
       \spp  { CE+GC+SMT }\spp   & 
     \spp  { CE+GC+NSC+PBH }\spp &
     \spp  { CE+GC+SMT+PBH }\spp 
        \\ 
{ w.o. GW190521}  &
1.22     &
        0.52     &
          1.39   &
        1.43     &
        1.31
\\
{ w. GW190521} &  
2.38   &
        -0.15    &
         0.72 &
        2.30 &
        2.58
           \\
 \hline
  \hline
\end{tabularx}
\label{tabbayes}
\end{table*}

\noindent{{\bf{\em Results.}}}
Among the events in the GWTC-2 catalog, we discard those with large false-alarm rate (GW190426, GW190719, GW190909)~\cite{Abbott:2020gyp} and two events involving neutron stars (GW170817, GW190425). GW190814~\cite{Abbott:2020khf} requires a separate treatment, since its secondary mass ($m_2\approx 2.6\msun$) would correspond  to either the lowest-mass astrophysical BH or to the highest-mass neutron star observed to date, challenging our current understanding of compact objects. For the moment we assume that the secondary component of GW190814 is a neutron star and neglect this event. 

Unlike Ref.~\cite{Zevin:2020gbd}, we do not exclude GW190521~\cite{Abbott:2020tfl}. At least the primary component of GW190521 lies in the (upper) mass gap predicted by pair-instability supernova theory, in tension with many astrophysical models (but see~\cite{Farmer:2019jed,Sakstein:2020axg,Belczynski:2020bca,Costa:2020xbc,Woosley:2021xba,Baxter:2021swn}), while being compatible with the PBH scenario~\cite{DeLuca:2020sae}. The selected catalog has $N_{\rm obs}=44$ or $43$, depending on whether GW190521 is included or not. 

\begin{figure*}[t!]
	\centering
	\includegraphics[width=0.49 \linewidth]{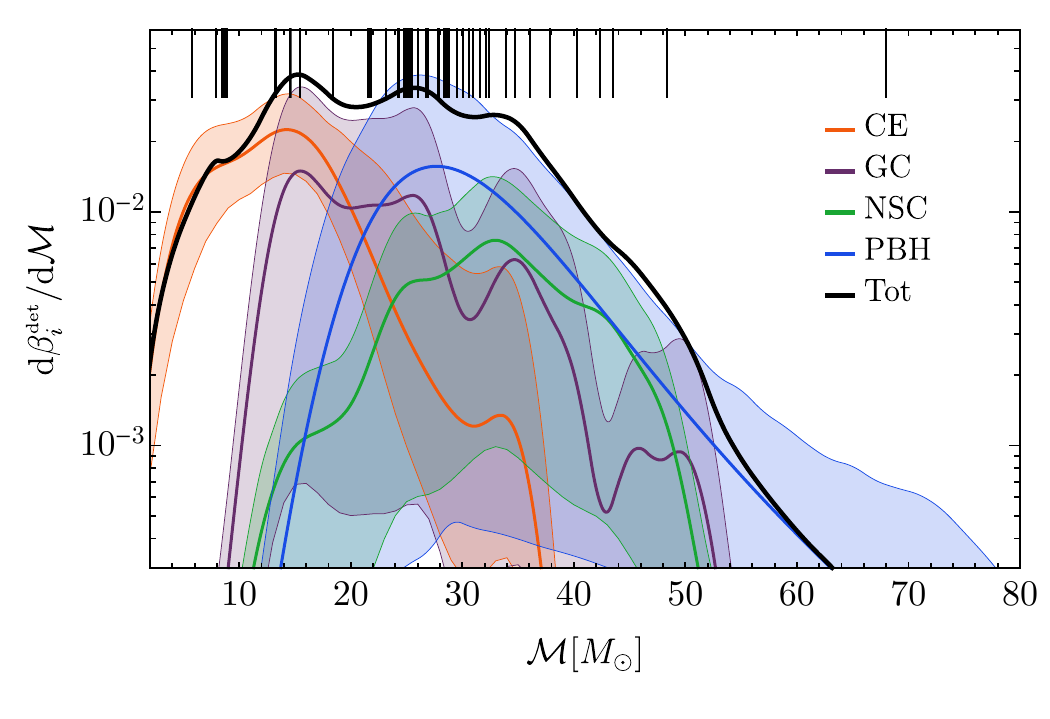}
\includegraphics[width=0.49 \linewidth]{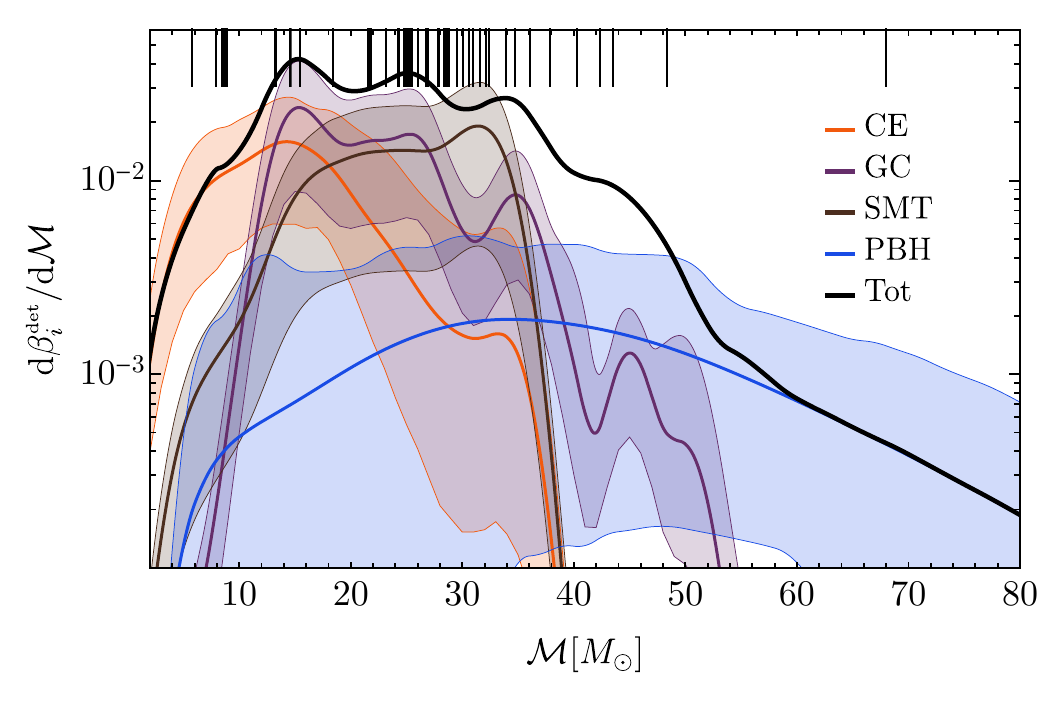}
\caption{
Observable distributions of chirp mass for each channel in the CE+GC+NSC+PBH (left) and CE+GC+SMT+PBH (right) scenario.
Analogous plots for other scenarios considered in the analysis are shown in the SM. The bands indicate $90\%$ C.I., while the black line corresponds to the mean total population. Vertical lines at the top of each plot correspond to the mean observed values for the events in the GWTC-2 catalog.}
	\label{fig:PDFseparate}
\end{figure*}

The results of our hierarchical Bayesian analysis are summarized in Fig.~\ref{fig:corner}, showing the posterior distributions of the detectable mixing fractions
$\beta_{i}^\text{\tiny det} = N^\text{\tiny det}_{i}/\sum_{j} N^\text{\tiny det}_{j}$, where ${i,j}=\{{\rm CE,GC,NSC,SMT,PBH}\}$ for the different models. We present various scenarios mixing the PBH population with different combinations of astrophysical channels: a simplified $2+1$ multichannel assuming only the two main astrophysical models (CE and GC, left panel), and two combinations of three astrophysical channels: CE+GC+NSC and CE+GC+SMT.
Table~\ref{tabbayes} shows the Bayes factors for various mixed scenarios with and without a PBH subpopulation. 

First of all, a two-channel CE+GC model is insufficient to explain GWTC-2 data. Either with or without GW190521, Fig.~\ref{fig:corner} (left panel) shows that in the CE+GC+PBH case the inferred PBH population fraction is approximately one third. Table~\ref{tabbayes} confirms that CE+GC+PBH is strongly favored over CE+GC, while the inclusion of NSC does not improve the overall fit. This is because the NSC and GC channels compete to explain similar events, whereas the PBH and GC channels produce different and complementary populations.

By comparing three-channel scenarios we see that models including NSC are not favored: NSCs account for some events in the central range of chirp masses, but the relative fraction of NSC events is small (both with and without GW190521).
The CE+GC+SMT channel has larger evidence, because the SMT channel complements CE and GC by predicting more massive binaries (but see e.g.~\cite{Olejak:2021fti} for a discussion of uncertainties in this prediction). However, even the best-fit SMT channel does not reach the mass gap. 

PBHs can efficiently produce binaries in the mass gap, so the inclusion of GW190521 leads to 
larger Bayesian evidence in favor of mixed astrophysical+PBH models. Furthermore,  PBHs can account for some of the heavy events other than GW190521 when the SMT channel is not included. This leads to a significant PBH fraction $\beta_\PBH^\text{\tiny det} =0.31^{+0.28}_{-0.26}\ (0.27^{+0.28}_{-0.24})$ at 90\% C.I. 
in the CE+GC+PBH (CE+GC+NSC+PBH) scenarios with GW190521, as shown in Fig.~\ref{fig:corner}. This conclusion would be unaffected even if GW190521 were an outlier belonging to a different astrophysical channel.

Let us now focus on four-channel scenarios. As shown in the insets of the right panel of Fig.~\ref{fig:corner}, when we include GW190521 the posterior of the PBH mixing fraction has vanishing support at $\beta_\PBH^\text{\tiny det}\approx0$.
The first percentile of $\beta_\PBH^\text{\tiny det}$ is ($0.022,0.014, 0.002$) for the (CE+GC+PBH, CE+GC+NSC+PBH, CE+GC+SMT+PBH) mixed scenarios.
The smallest PBH fraction ($\beta_\PBH^\text{\tiny det} =0.06^{+0.15}_{-0.05}$) corresponds to CE+GC+SMT+PBH: the SMT channel can reproduce most events below the mass gap, and only GW190521 is confidently interpreted as a PBH binary. 
If we exclude GW190521 from the catalog, the posterior distribution of $\beta_\PBH^\text{\tiny det}$ ``flattens out'' in the CE+GC+SMT+PBH scenario, becoming compatible with zero (blue histogram in the top-right inset of Fig.~\ref{fig:corner}). This conservative scenario suggests that the PBH fraction can be compatible with zero if the mass gap event is interpreted in other ways. For example,  
under different assumptions on the prior, the event could be interpreted as a   straddling binary (see e.g. \cite{Fishbach:2020qag,Nitz:2020mga,Estelles:2021jnz}).
Furthermore, heavy binaries like GW190521 could form in AGN disks~\cite{McKernan:2019beu,Tagawa:2019osr,Yang:2019okq}.
On the other hand, as shown in the SM, $\beta_\PBH^\text{\tiny det}$ does not depend significantly on the inclusion of the lower mass-gap event GW190814.

To better understand how the GWTC-2 events are interpreted by the inference, in Fig.~\ref{fig:PDFseparate} we plot the contribution of each population to the observed chirp mass distribution for the CE+GC+NSC+PBH model (left) and for the CE+GC+SMT+PBH model (right). 
The PBH population overlaps mostly with the GC channel (and with NSC/SMT, when included in the inference), but it recovers larger values of ${\cal M}$. 
As the PBH distribution extends to reach GW190521, it becomes less competitive at explaining the bulk of events in the central range of ${\cal M}$. 
For this reason the posterior of $f_\PBH$ has a significant tail at small values (see Fig.~\ref{fig:corner_combined} in the SM).
This does not happen when GW190521 is removed, since then the PBH model can efficiently reproduce events in the central range of ${\cal M}$.
Note also that most low-${\cal M}$ events come from the CE channel in all cases.

\noindent{{\bf{\em Discussion.}}}
We have presented a robust analysis aimed at constraining the relative fraction of different BBH formation channels, including both state-of-the-art astrophysical populations~\cite{Zevin:2020gbd} and a subpopulation of PBHs. Our main result is that the relative PBH abundance in GWTC-2 data depends on the astrophysical channels included in the analysis. In particular, while a PBH population is statistically favored against competitive astrophysical models, such as GC and NSC, 
a dominant contribution from the SMT channel along with CE and GC drastically reduces the need for PBHs, except for explaining the rate of mass-gap events like GW190521. If we further exclude GW190521, the Bayesian evidence for CE+GC+SMT becomes comparable to CE+GC+SMT+PBH, showing that the constraining power of the current data set is not sufficient to draw firm conclusions.
The fraction of SMT events necessary to explain the data is $0.34\pm0.22$, so that SMT would have to be the dominant channel. 
Our main results would not be qualitatively affected by including recently reported events~\cite{LIGOScientific:2021djp}, since the bulk mass and spin distribution of the GWTC-3 catalog is consistent with GWTC-2~\cite{LIGOScientific:2021psn}.

Overall, we conclude that confidently constraining a putative primordial population in GW data requires a more robust understanding of astrophysical populations. We have considered four state-of-the-art astrophysical models~\cite{Zevin:2020gbd}, but each of them is affected by large uncertainties, and there might exist others which are competitive against the primordial subpopulation.
On the PBH side, we adopted a standard lognormal distribution for the PBH mass function at formation, but it would be important to test the impact of other (model-dependent) assumptions and of different priors on the PBH hyperparameters motivated by specific formation mechanisms (see e.g.~\cite{Gow:2020cou,Gow:2020bzo}).

Confidently claiming the primordial nature of some individual BBHs would probably require single-event analyses for large signal-to-noise ratio events, especially by cross-correlating merger rates with mass, spin, and redshift measurements to identify key features of the PBH scenario (see e.g. \cite{Bhagwat:2020bzh}). 
Another possibility to break the degeneracy between the PBH and astrophysical channels is to perform population studies focusing on spin distributions~\cite{Fernandez:2019kyb,Garcia-Bellido:2020pwq}, and accounting for the $q-\chi_\text{\tiny eff}$ correlation introduced by accretion effects in PBH models~\cite{DeLuca:2020bjf,DeLuca:2020qqa}.

A conclusive verdict on the primordial nature of a subpopulation of BBHs  may come from third-generation GW detectors such as the Einstein Telescope~\cite{Hild:2010id} and Cosmic Explorer~\cite{Dwyer:2014fpa}, that will detect BH mergers up to $z\approx 50$~\cite{Maggiore:2019uih}, and can in principle reconstruct the redshift evolution of the merger rate (although the accuracy of the redshift measurement deteriorates 
with redshift
\cite{Vitale:2016icu,Ng:2020qpk}). 
The merger rate is monotonically increasing with redshift for primordial BBHs, whereas it should peak around $z\approx 2$ for astrophysical BBHs, and at $z\approx 10-20$ for BHs formed from the first stars~\cite{Schneider:1999us,Schneider:2001bu,Schneider:2003em} (see~\cite{Hartwig:2016nde,Belczynski:2016ieo,Vitale:2018yhm,Ng:2020qpk,Liu:2020ufc,Valiante:2020zhj} for recent studies).
Note that a fraction $\beta^\text{\tiny det}_\PBH={\cal O}(10\%)$ in current data would be in agreement with the simplified analysis of Ref.~\cite{DeLuca:2021wjr} using the LVC power-law model for the astrophysical population. By mapping this fraction of PBHs to the merger rates for third-generation detectors, one would expect dozens to hundreds BBH detections at $z\gtrsim 30$, which might be identified as primordial~\cite{Koushiappas:2017kqm,DeLuca:2021wjr}, as long as we can accurately measure their redshift~\cite{Ng:2020qpk}.
Alternatively, another test of the presence of PBHs in GW data may come from a population analysis of the events measurable at high redshift by third-generation interferometers, thus exploiting the information on the merger rate evolution~\cite{inprep}.

\begin{table*}[t]
\renewcommand{\arraystretch}{1.3}
\caption{Prior ranges for the hyperparameters of the primordial and astrophysical models. We assume uniform distributions for all parameters.
In particular, we adopt a flat prior in $\log(f_\text{\tiny PBH})$ for the PBH abundance.
Also, the prior on $M_c$ was extended up to $50 M_\odot$ 
in the scenario including the SMT channel, because the PBH population shifts to larger masses.
Following Ref.~\cite{Zevin:2020gbd}, we considered discrete values for $\alpha_\text{\tiny \rm CE}$ and $\chi_{\rm b}$. Binary components spinning at $\chi<\chi_{\rm b}$ at BBH formation are given spins of $\chi_{\rm b}$.
We recall that $\alpha_\text{\tiny\rm CE}$ only affects the CE model.}
\begin{tabularx}{1.95\columnwidth}{X|cccc|cc}
\hline
  \hline
  Model &
  \multicolumn{4}{c|}{PBH}   &
  \multicolumn{2}{c}{ABH} \\
  \hline
Parameter&     
      \hspace{.3cm}  $M_c\, [M_\odot]$ \hspace{.3cm}  &
      \hspace{.3cm}    $\sigma$ \hspace{.3cm} &
   \hspace{.3cm}      $\log f_\text{\tiny PBH}$  \hspace{.3cm}  & 
   \hspace{.3cm}   $z_\text{\tiny cut-off}$   \hspace{.3cm}  & 
         \hspace{.8cm} $\alpha_\text{\tiny CE}$ \hspace{.8cm} &
      \hspace{.8cm}   $\chi_\text{\tiny b}$ \hspace{.8cm}  \\ 
Prior range &  $[10,40]$ & $[0.1,1.1]$  & $[-5,-2]$ & $[10,30]$  & 
           [0.2,0.5,1,2,5] &
           [0,0.1,0.2,0.5] \\
 \hline
  \hline
\end{tabularx}
\label{tab:priors}
\end{table*}

\noindent{{\bf{\em Acknowledgments.}}}
We are very grateful to M.~Zevin for insightful discussions about Ref.~\cite{Zevin:2020gbd} and help in using the publicly available repository \cite{michael_zevin_2021_4448170}.
We also thank S.~Bavera for explaining some features of the SMT channel, and K.~Belczynski, C.~Byrnes, D.~Gerosa, A.~Gow, A.~Hall, D.~Holz, and M.~Mapelli for discussions.
Some computations were performed at the University of Geneva on the Baobab/Yggdrasil cluster and at Sapienza University of Rome on the Vera cluster of the Amaldi Research Center funded by the MIUR program ``Dipartimento di Eccellenza''~(CUP: B81I18001170001).
E.~Berti, V.~Baibhav and K. W. K. Wong are supported by NSF Grants No. PHY-1912550 and AST-2006538, NASA ATP Grants No. 17-ATP17-0225 and 19-ATP19-0051, NSF-XSEDE Grant No. PHY-090003, and NSF Grant PHY-20043. This work has received funding from the European Union’s Horizon 2020 research and innovation programme under the Marie Skłodowska-Curie grant agreement No. 690904. This research project was conducted using computational resources at the Maryland Advanced Research Computing Center (MARCC). 
V.DL., G.F. and A.R. are supported by the Swiss National Science Foundation 
(SNSF), project {\sl The Non-Gaussian Universe and Cosmological Symmetries}, project number: 200020-178787.
P.P. acknowledges financial support provided under the European Union's H2020 ERC, Starting Grant agreement no.~DarkGRA--757480, under the MIUR PRIN and FARE programmes (GW-NEXT, CUP:~B84I20000100001).
K. K. Y. N. and S. V., members of the LIGO Laboratory, acknowledge the support of the National Science Foundation through the NSF Grant No.
PHY-1836814. LIGO was constructed by the California Institute of Technology and Massachusetts Institute of Technology with funding from the National Science Foundation and operates under Cooperative Agreement No. PHY-1764464
The authors would like to acknowledge networking support by the GWverse COST Action CA16104, ``Black holes, gravitational waves and fundamental physics.''

This research has made use of data, software and/or web tools obtained from the Gravitational Wave Open Science Center (https://www.gw-openscience.org/ ), a service of LIGO Laboratory, the LIGO Scientific Collaboration and the Virgo Collaboration. LIGO Laboratory and Advanced LIGO are funded by the United States National Science Foundation (NSF) as well as the Science and Technology Facilities Council (STFC) of the United Kingdom, the Max-Planck-Society (MPS), and the State of Niedersachsen/Germany for support of the construction of Advanced LIGO and construction and operation of the GEO600 detector. Additional support for Advanced LIGO was provided by the Australian Research Council. Virgo is funded, through the European Gravitational Observatory (EGO), by the French Centre National de Recherche Scientifique (CNRS), the Italian Istituto Nazionale di Fisica Nucleare (INFN) and the Dutch Nikhef, with contributions by institutions from Belgium, Germany, Greece, Hungary, Ireland, Japan, Monaco, Poland, Portugal, Spain.
\appendix

\section{Supplemental material}

Here we provide details on our data analysis and further results complementary to those presented in the main text.

\subsection{Details of the data analysis}

The key quantity to be evaluated is the likelihood function $p(\bm{\lambda}|\bm{d})$, defined in Eq.~(1) of the main text. Here we explain how its various ingredients are computed.

The priors on the hyperparameters $\pi(\bm{\lambda})$ are uniformly distributed in the ranges given in Table~\ref{tab:priors}. Note that at values of the cut-off redshift above  $z_\co\gtrsim30$, and in the mass range of interest,
accretion onto PBHs is negligible~\cite{DeLuca:2020qqa}. 
Therefore, all models with $z_\co\gtrsim30$
are degenerate and we can cut the range at this reference value, as done in Refs.~\cite{Wong:2020yig,DeLuca:2021wjr}.

\begin{figure*}[th]
\centering
\includegraphics[width=0.492\textwidth]{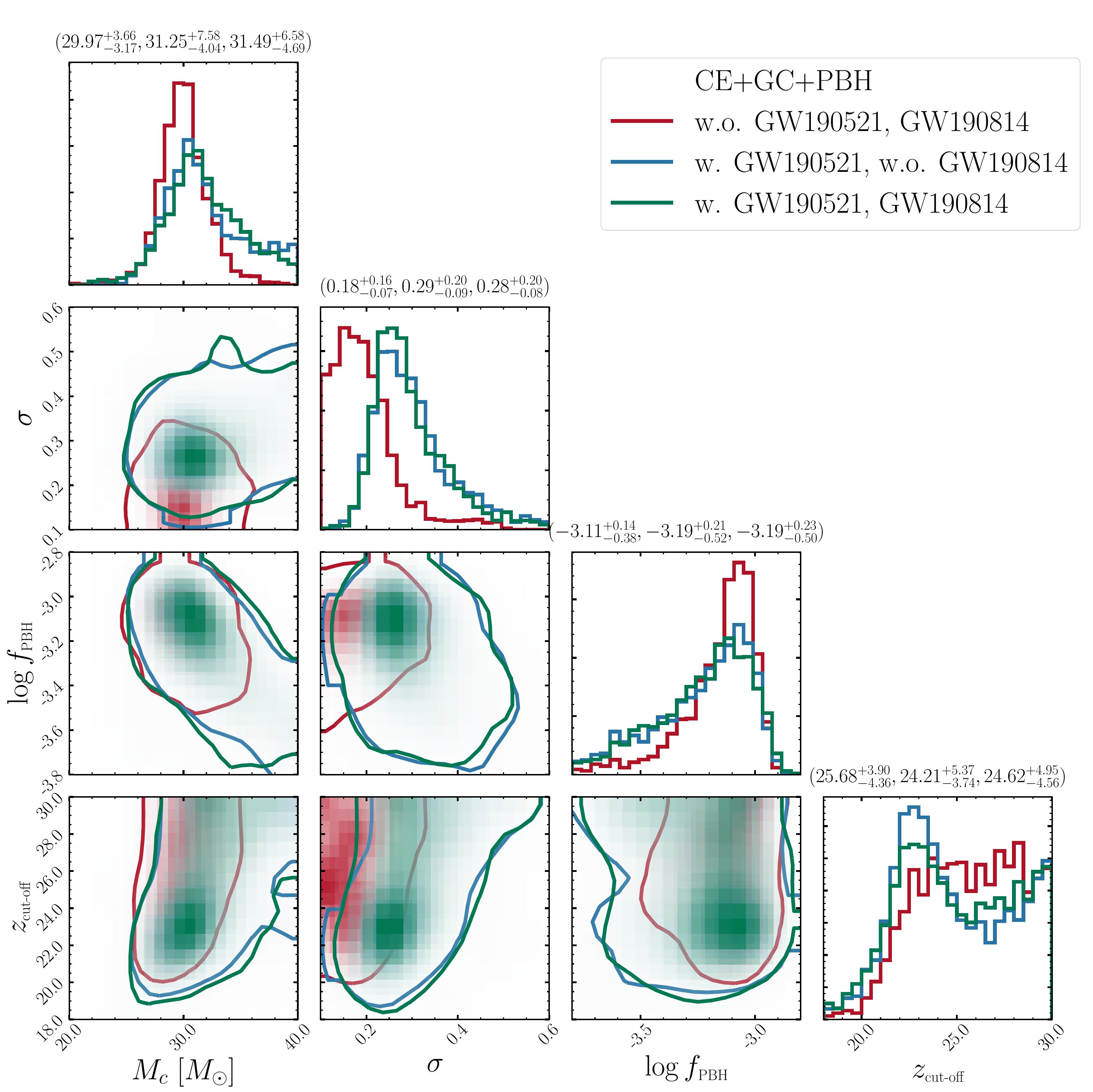}
\includegraphics[width=0.488\textwidth]{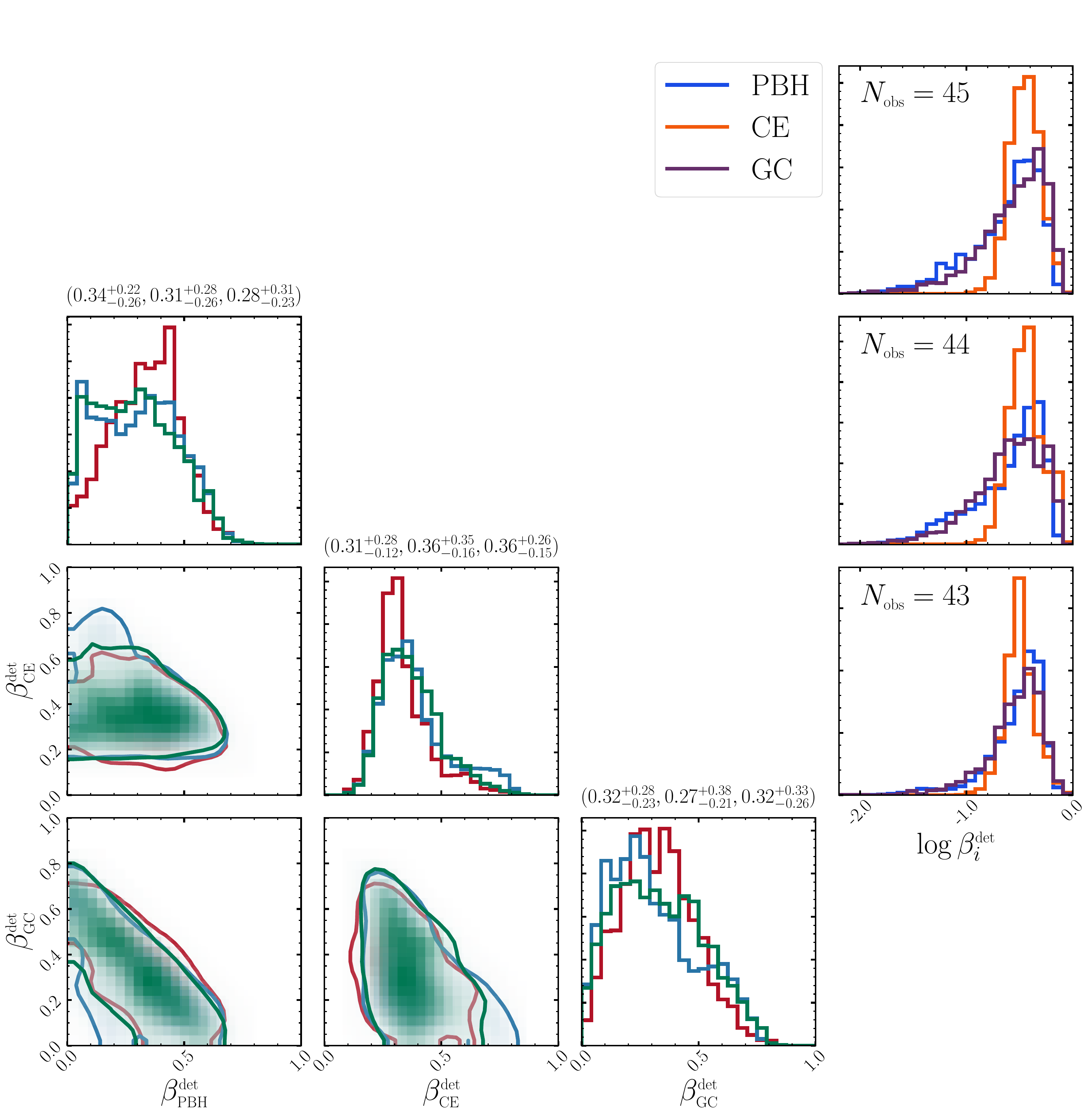}
\caption{
Posterior distributions of the hyperparameters of the PBH model (left panel) and of the individual {\it detectable} mixing fractions $\beta_{i}^\text{\tiny det}$ of different populations (right panel) for the  CE+GC+PBH model.
Each color corresponds to the case both GW190521 and GW190814 are neglected (red, $N_\text{\tiny obs}=43$), only GW190521 is additionally included (blue, $N_\text{\tiny obs}=44$) and
both GW190521 and GW190814 are included (green, $N_\text{\tiny obs}=45$).
The $90\%$ C.I. reported on top of each column correspond to the various cases, following the (red, blue, green) ordering. 
 The insets in the top right
 show the individual $\beta_{i}^\text{\tiny det}$ on a logarithmic scale.
}\label{fig:corner_combined}
\end{figure*}

The binary parameter distributions in a given model (either primordial or astrophysical) can be computed from the differential merger rate $\d R/( \d m_1 \d m_2)$ as 
\begin{equation}
p_\text{\tiny pop} (\bm{\theta}|\bm{\lambda}) \equiv \frac{1}{N({\bm \lambda})} \llp  T_\text{\tiny obs} \frac{1}{1+z} \frac{\d V}{\d z} \frac{\d R}{\d m_1 \d m_2} (\bm{\theta}|\bm{\lambda})\rrp,
\label{ppop}
\end{equation}
with $T_\text{\tiny obs}$ being the observation time,  
whereas the number of expected detections reads
\begin{align}
N_\text{\tiny det}(\bm{\lambda})
 &\equiv 
  T_\text{\tiny obs}
   \int \d m_1 \d m_2 \d z\, p_\text{\tiny det} (m_1, m_2, z) 
   \nonumber \\
  & \times  \frac{1}{1+z} \frac{\d V}{\d z} \
  \frac{\d R}{\d m_1 \d m_2}  (m_1, m_2, z|\bm{\lambda}) \,,
\end{align}
where the prefactor $1/(1+z)$ accounts for the redshift at the source epoch and $\d V / \d z$ stands for the differential comoving volume factor, see e.g.~\cite{Dominik:2014yma}.
We account for selection bias by introducing the probability of detection  
\begin{equation}
p_\text{\tiny det}(\theta_\text{\tiny i})= \int  p(\theta_\text{\tiny e}) \, \Theta[\rho(\theta_\text{\tiny i},\theta_\text{\tiny e}) - \rho_\text{\tiny  thr}] \, d \theta_\text{\tiny e}\,,
\label{pdetdefine}
\end{equation}
where $\theta_\text{\tiny i}=\{m_1,m_2,z\}$ are the intrinsic parameters of the binary (individual source-frame masses $m_i$ and merger redshift $z$), whereas $\theta_\text{\tiny e} = \{\alpha,\delta,\iota,\psi\}$ are the extrinsic parameters (right ascension $\alpha$, declination $\delta$, orbital-plane inclination $\iota$, and polarization angle $\psi$). Finally, $p(\theta_\text{\tiny e})$ is the probability distribution function of $\theta_\text{\tiny e}$, $\Theta$ is the Heaviside step function, and $\rho$ is the signal-to-noise ratio~(SNR). 
For simplicity we neglect the spins $\boldsymbol{\chi}_i$ ($i=1,2$) in the computation of the detectability, since the large majority of the GWTC-2 events are compatible with zero spin.

In the case of the GWTC-1 catalog, $p_\text{\tiny det}$ can be computed in the single-detector semianalytic framework of Refs.~\cite{Finn:1992xs,Finn:1995ah} and adopting a SNR threshold $\rho_\text{\tiny thr}=8$ without encountering significant departures from the large-scale injection campaigns in the O1~\cite{Abbott:2016nhf} and O2~\cite{LIGOScientific:2018jsj} runs. We adopt the same procedure to compute the detectability of binaries also for the O3a run.

The SNR can be factored out as  $\rho(\theta_\text{\tiny i},\theta_\text{\tiny e})=\omega(\theta_\text{\tiny e}) \rho_\text{\tiny opt}(\theta_\text{\tiny i})$, where $\rho_\text{\tiny opt}$ is the SNR of an ``optimal'' source located overhead the detector with face-on inclination. 
The optimal SNR $\rho_\text{\tiny opt}$ of individual GW events is given in terms of the (Fourier-transformed) GW waveform by
\begin{equation}
\rho_\text{\tiny opt}^2 (m_1,m_2,\boldsymbol{\chi}_1,\boldsymbol{\chi}_2,z)\equiv \int_0^\infty \frac{4 |\tilde h (\nu)|^2}{S_n (\nu)} {\rm d} \nu,
\end{equation}
where $S_n$ is the strain noise of the detector. 
Following the choice adopted in Ref.~\cite{Zevin:2020gbd}, we adopt the
 \texttt{midhighlatelow} noise power spectral densities \cite{Aasi:2013wya}, as implemented in the publicly available repository \texttt{pycbc}~\cite{alex_nitz_2021_4556907}.

\begin{figure*}[th]
\centering
\includegraphics[width=0.49\textwidth]{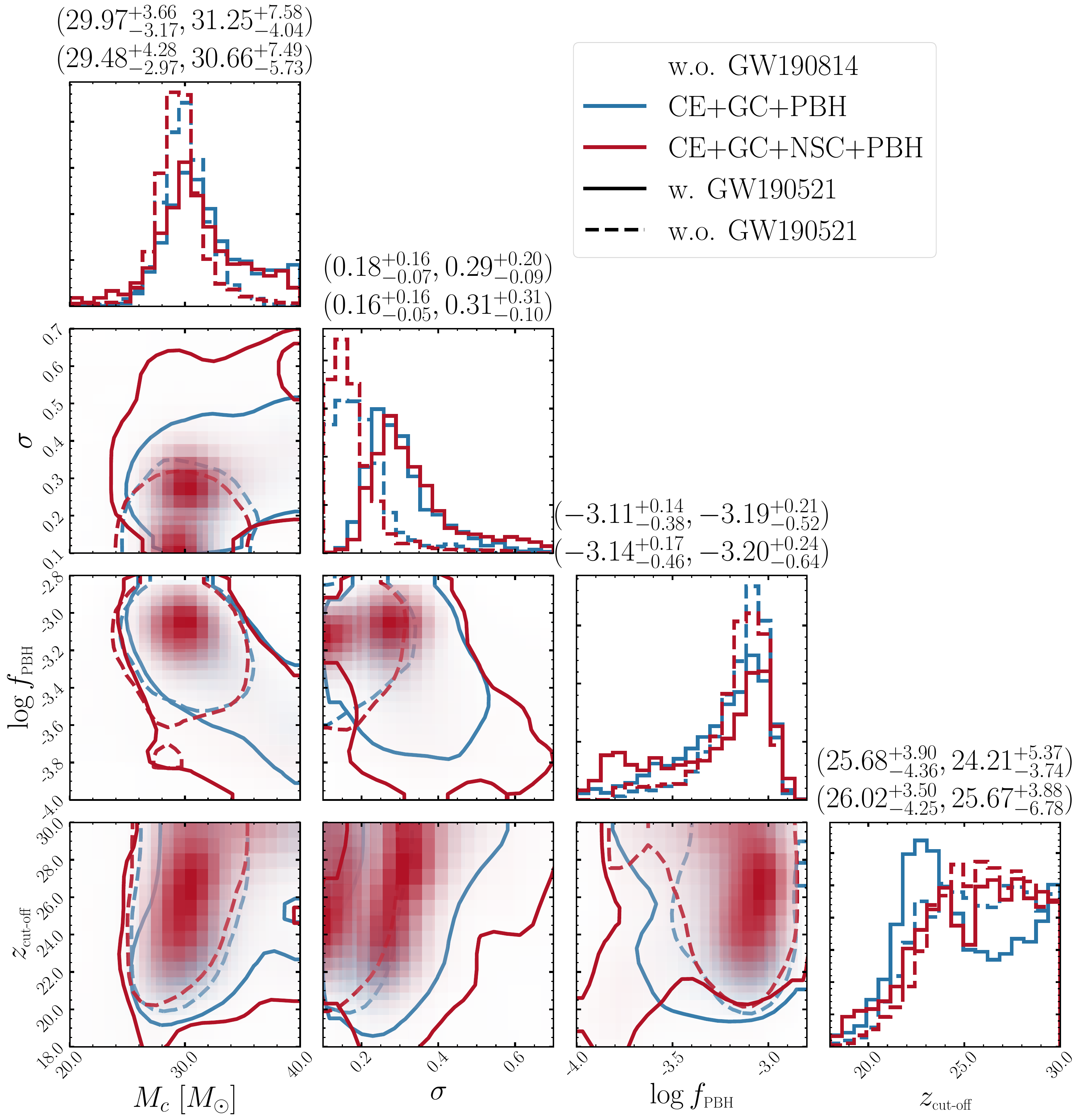}
\includegraphics[width=0.49\textwidth]{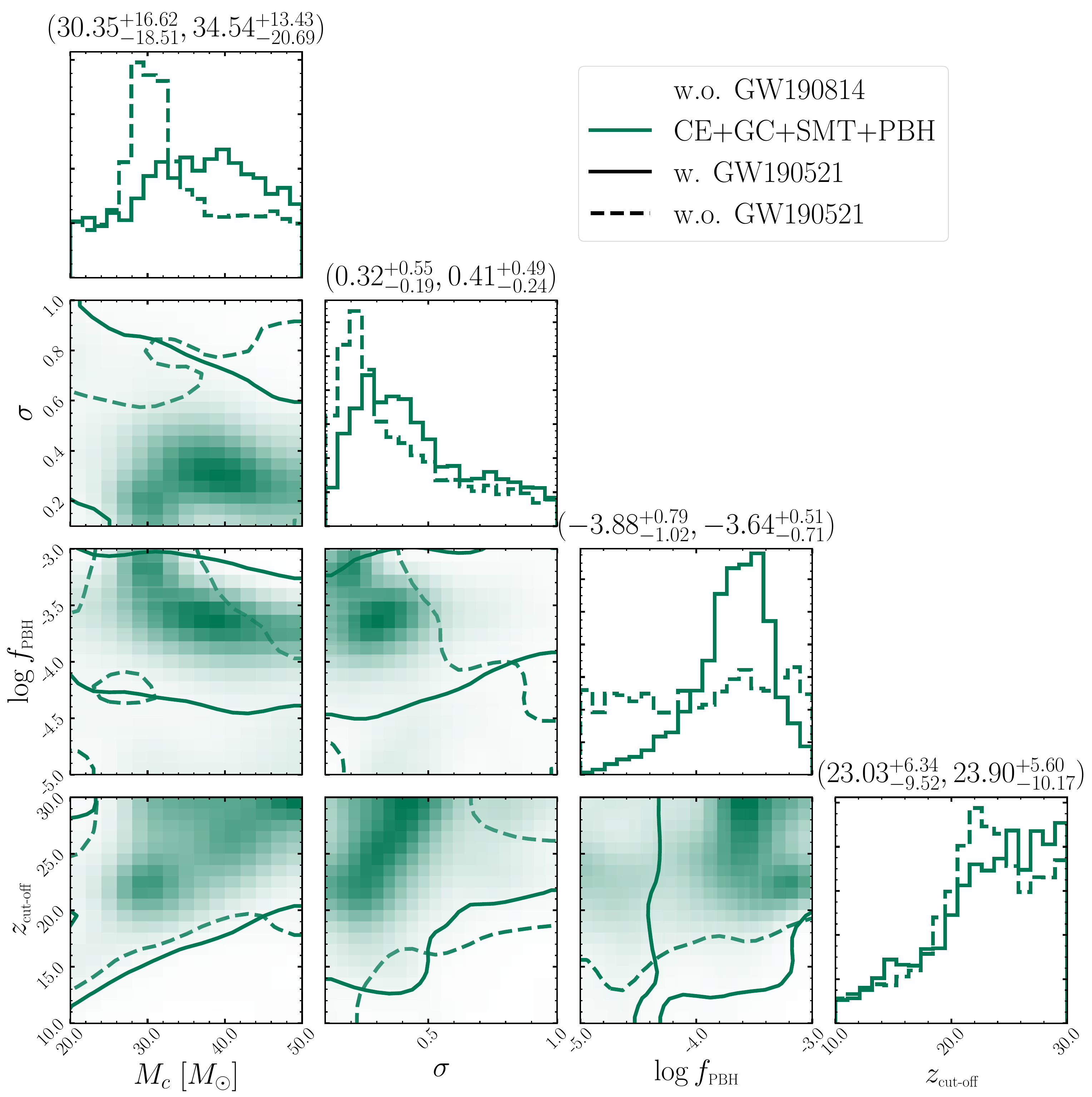}
\caption{
 Posterior distributions of the hyperparameters of the PBH model  for the  CE+GC+PBH, CE+GC+NSC+PBH and CE+GC+SMT+PBH mixed models (blue, red and green, respectively) with (solid line) and without (dashed line) the mass gap event GW190521, and excluding GW190814. 
 The left (right) values of the 90\% C.I. refer to the cases without (with) GW190521. 
}\label{fig:2p1_3p1_noGW190521}
\end{figure*}

Finally, we explicitly compute the marginalized distribution $p_\text{\tiny det}(\theta_\text{\tiny i})$ [Eq.~\eqref{pdetdefine}] by evaluating the cumulative distribution function $P(\omega_\text{\tiny thr})=\int_{\omega_\text{\tiny thr}}^1 p(\omega') d\omega'$ at  $\omega_\text{\tiny thr}=\rho_\text{\tiny thr}/\rho_\text{\tiny opt}(\theta_\text{\tiny i})$. We consider isotropic sources, so that $\alpha, \cos\delta, \cos\iota$, and $\psi$ are uniformly distributed. Then, for the case of a single-detector approximation, nonprecessing binaries, and considering only the dominant quadrupolar mode, the function $P(\omega_\text{\tiny thr})$ is found as in Ref.~\cite{Dominik:2014yma}.

As discussed in the main text, we use the GWTC-2 catalog~\cite{Abbott:2020gyp}, discarding three events with large false-alarm rate (GW190426, GW190719, GW190909) and two events involving neutron stars (GW170817, GW190425). The cases of GW190814~\cite{Abbott:2020khf} and GW190521~\cite{Abbott:2020tfl} require special treatment. The former can be a BH-neutron star binary with the heaviest neutron star to date, whereas the latter is in tension with the main astrophysical formation channels we adopt, since its primary lies within the pair-instability mass gap predicted by supernova theory.
In the following, in order to remain agnostic about the nature of GW190814, we shall present some results of our analysis both with and without this event, showing that its inclusion does not alter our results as it would be interpreted in all cases as a binary coming from the CE channel.
For a given set of astrophysical populations models, the mass-gap event play a crucial role in determining the inferred fraction of PBHs in the catalog. For this reason we performed the inference both with and without GW190521. 
Overall, the selected catalog contains $43$ events + GW190521 + GW190814, hence $N_\text{\tiny obs}=43,44,45$, depending on the setup.
Similarly to Ref.~\cite{Zevin:2020gbd}, we adopt the ``Combined'' samples for the GWTC-1 events as provided in~\cite{119}, and the ``PublicationSamples'' in~\cite{120} for the GWTC-2 events.

We compute the evidence for each model from the posterior data following Ref.~\cite{NR1994}. We do not expect the error associated with the estimation of the evidence to affect our conclusions. For example, we repeated the computation of the evidence for the CE+GC+PBH scenario adopting a nested sampling algorithm, as implemented in DYNESTY \cite{Speagle:2019ivv}, finding a result consistent with Table~\ref{tabbayes} and an error of $\Delta (\log_{10} Z) = \pm  0.2$ (90\% C.I.).

\subsection{Supplemental results}

First of all, we checked the robustness of our results against the inclusion of the asymmetric merger GW190814.
In Fig.~\ref{fig:corner_combined}, we compare the posterior distributions obtained including/excluding GW190521 and GW190814 in various combinations for the CE+GC+PBH mixed model. 
The inclusion of GW190814 has a mild effect: this event is always ascribed to the CE population, since the latter has the strongest support at small masses. 

In Fig.~\ref{fig:2p1_3p1_noGW190521} we show the posterior distributions of the hyperparameters of the PBH model for the CE+GC+PBH, CE+GC+NSC+PBH and CE+GC+SMT+PBH mixed scenarios, both with and without GW190521. These results complement those shown in Fig.~\ref{fig:corner}, where we showed the corresponding observable mixing fractions. 

The posterior distributions in the first two cases (CE+GC+PBH and CE+GC+NSC+PBH) are strikingly similar. The only relevant differences are found when comparing results with and without GW190521.
When the mass gap event is included, we find a larger width $\sigma$ of the PBH mass function and a slightly enhanced tail of the posterior distribution of $M_c$ at large values, because the PBH channel is necessary to produce heavy binaries in the mass gap. 

In the CE+GC+PBH mixed model including the mass gap event, the distribution of $z_\text{\tiny cut-off}$ shows two peaks at $z_\text{\tiny cut-off}\approx 23$ and $z_\text{\tiny cut-off}\approx 30$.
The first corresponds to the case where some PBH accretion is necessary to explain the (few) spinning events in the catalog~\cite{DeLuca:2020bjf, DeLuca:2020qqa}, while the second peak corresponds to the case where the observed events associated to PBHs by the inference are mostly nonspinning.
In the CE+GC+NSC+PBH mixed case, the posterior of $z_\text{\tiny cut-off}$ is approximately flat above $z_\text{\tiny cut-off}\sim 25$, which is possibly explained by the fact that fewer events with nonnegligible spin are assigned to PBHs. We have also checked that the posterior remains flat for $z_\text{\tiny cut-off}\gsim{30}$, where accretion is indeed negligible in the mass range of interest.

\begin{figure*}[t!]
	\centering
	\includegraphics[width=0.475 \linewidth]{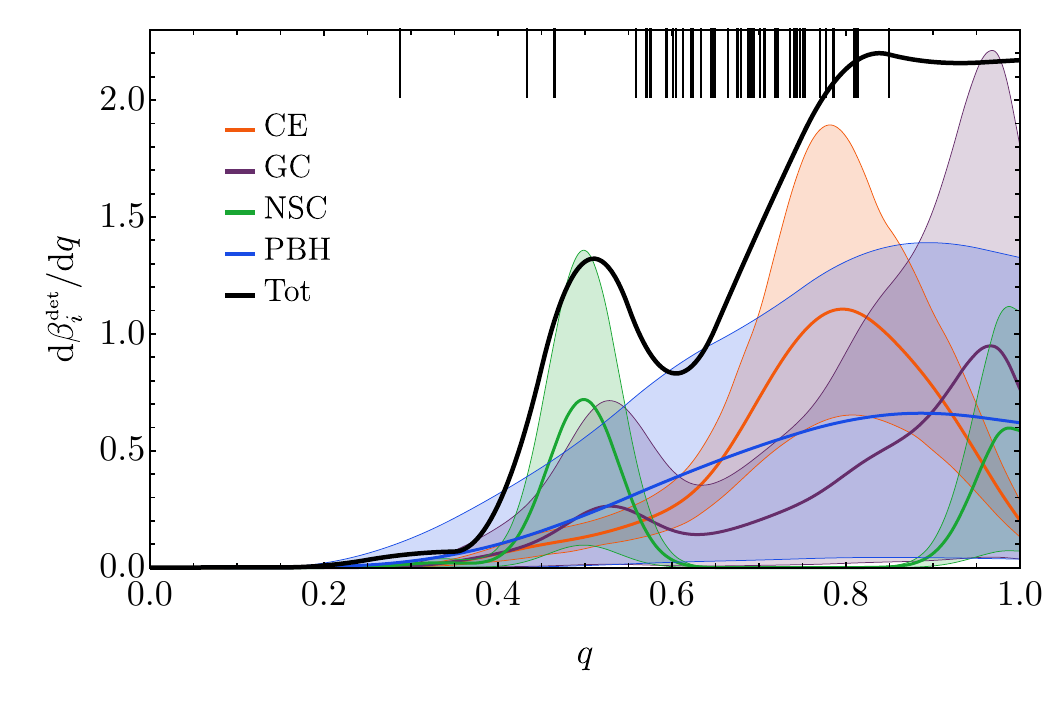}
\includegraphics[width=0.475 \linewidth]{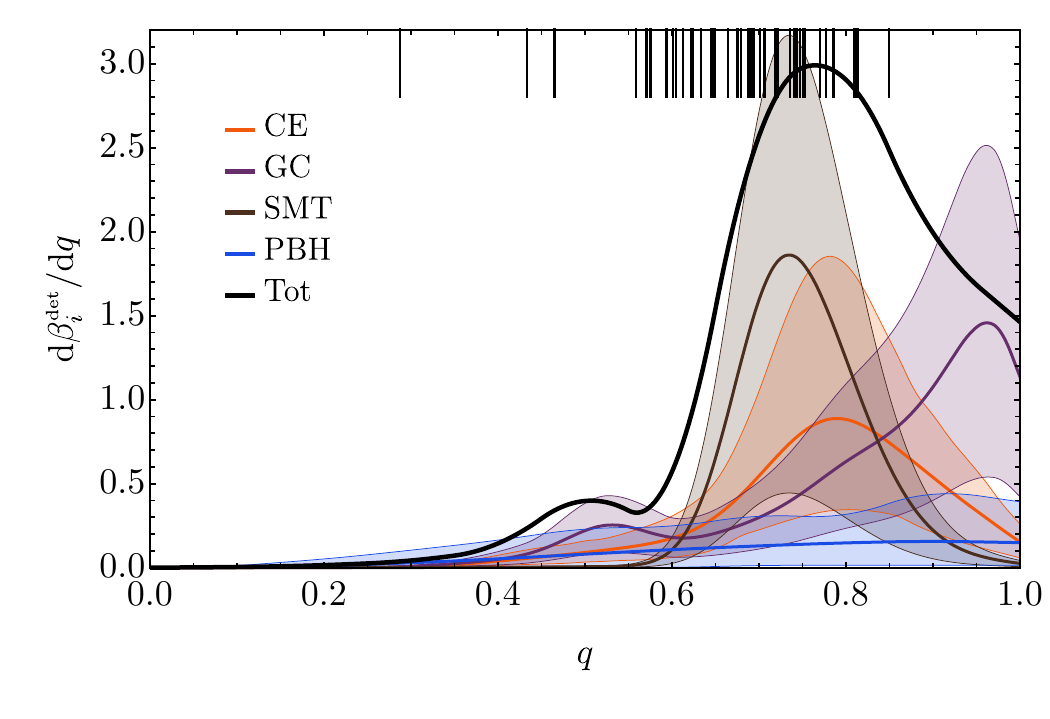}
\caption{
Mass ratio distribution for the CE+GC+NSC+PBH (left) and CE+GC+SMT+PBH (right) mixed scenario including GW190521. This figure complements the chirp mass distributions shown in Fig.~\ref{fig:PDFseparate}.}
	\label{fig:PDFseparate_5}
\end{figure*}

\begin{figure*}[t!]
	\centering
	\begin{subfigure}[b]{1\textwidth}
          \centering
	\includegraphics[width=0.475 \linewidth]{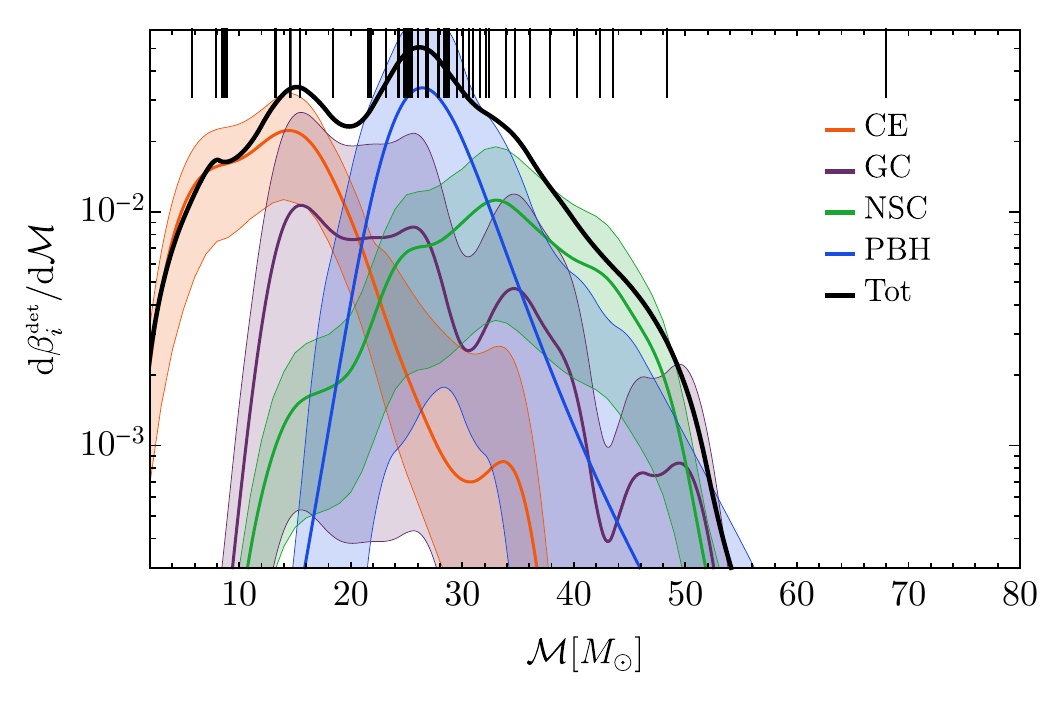}
\includegraphics[width=0.475 \linewidth]{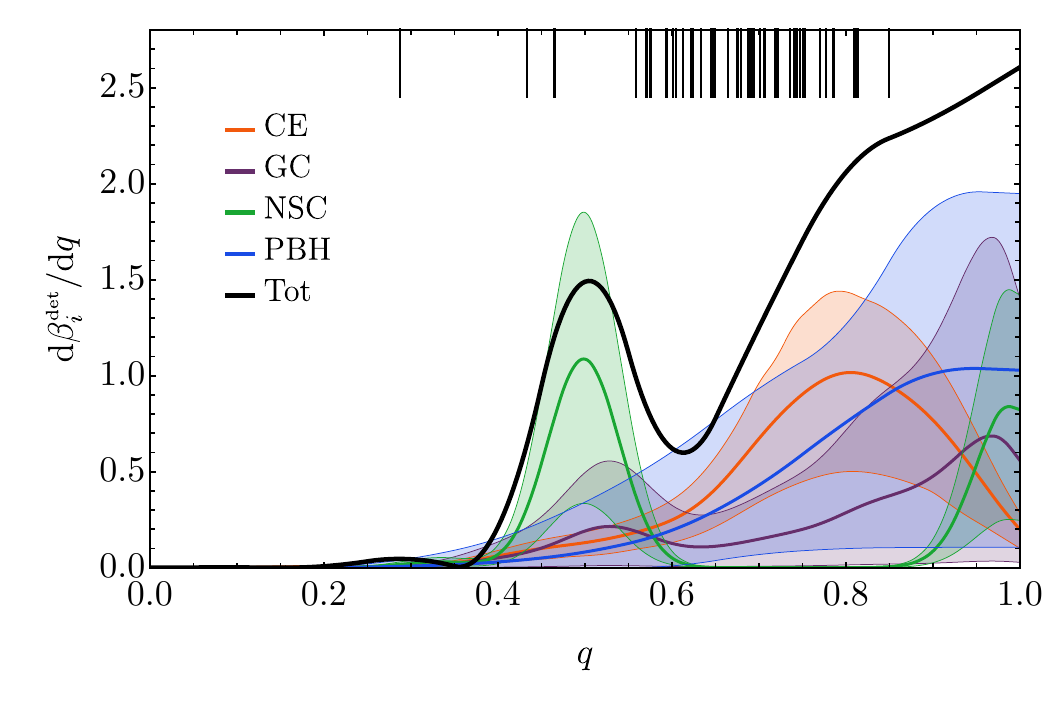}
\captionsetup{justification=centering, singlelinecheck=false}
	\caption{ CE+GC+NSC+PBH mixed scenario without including GW190521.}
	\label{fig:PDFseparate_6a}
	\end{subfigure}
	\begin{subfigure}[b]{1\textwidth}
          \centering
	\includegraphics[width=0.475 \linewidth]{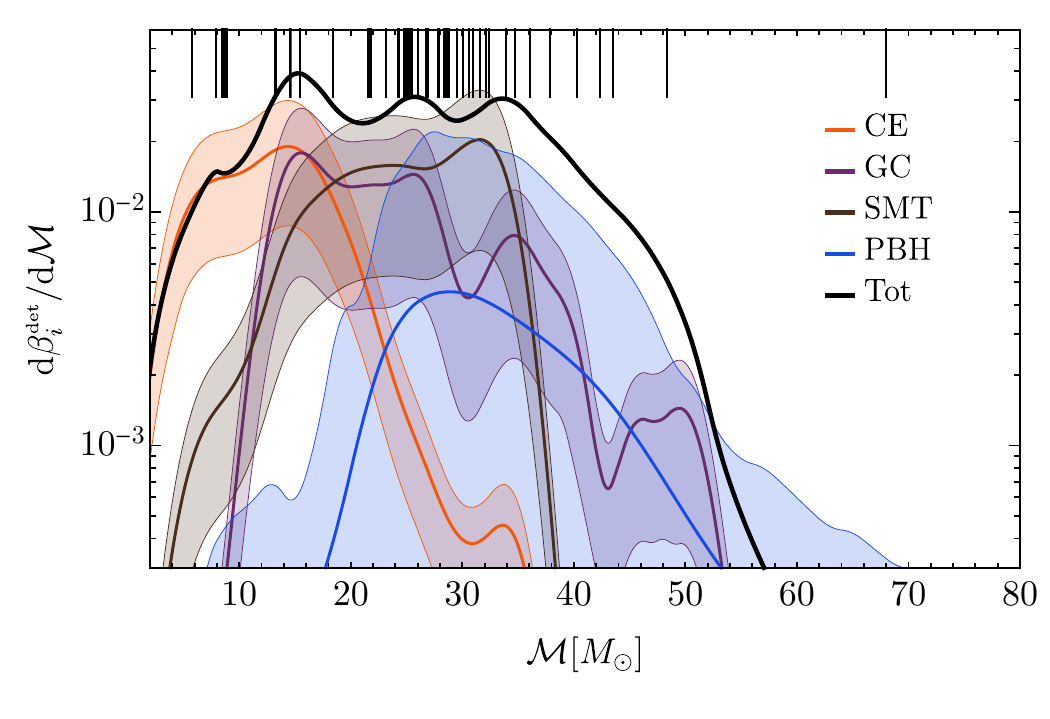}
\includegraphics[width=0.475 \linewidth]{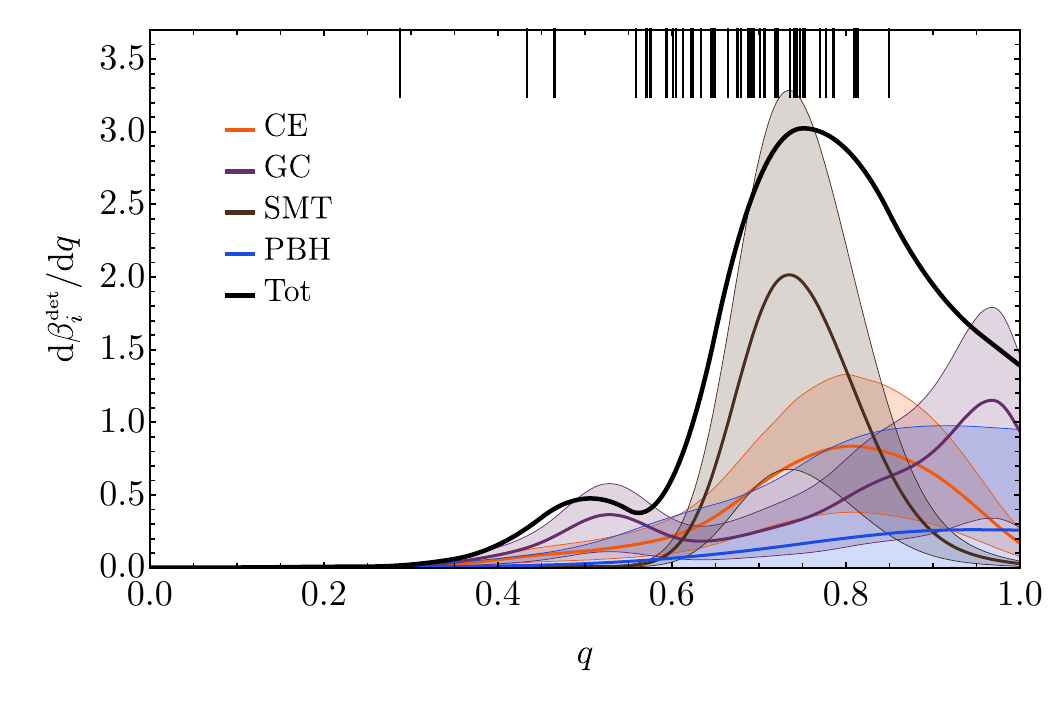}
\captionsetup{justification=centering, singlelinecheck=false}
	\caption{ CE+GC+SMT+PBH mixed scenario without including GW190521.}
	\label{fig:PDFseparate_6b}
	\end{subfigure}
	\caption{
	Individual contribution to the observable distributions of chirp mass (left) and mass ratio (right) for 
	each 3+1 scenarios without the mass gap event GW190521.
In the bottom panels, the PBH contribution is unbounded from below, being the observable fraction $\beta_\PBH^\text{\tiny det}$, in this analysis without including GW190521 and with SMT as a sub-population, compatible with zero.  
}
 	\label{fig:PDFseparate_6}
\end{figure*}

\begin{figure*}[t!]
	\centering
	\begin{subfigure}[b]{1\textwidth}
          \centering
		\includegraphics[width=0.475 \linewidth]{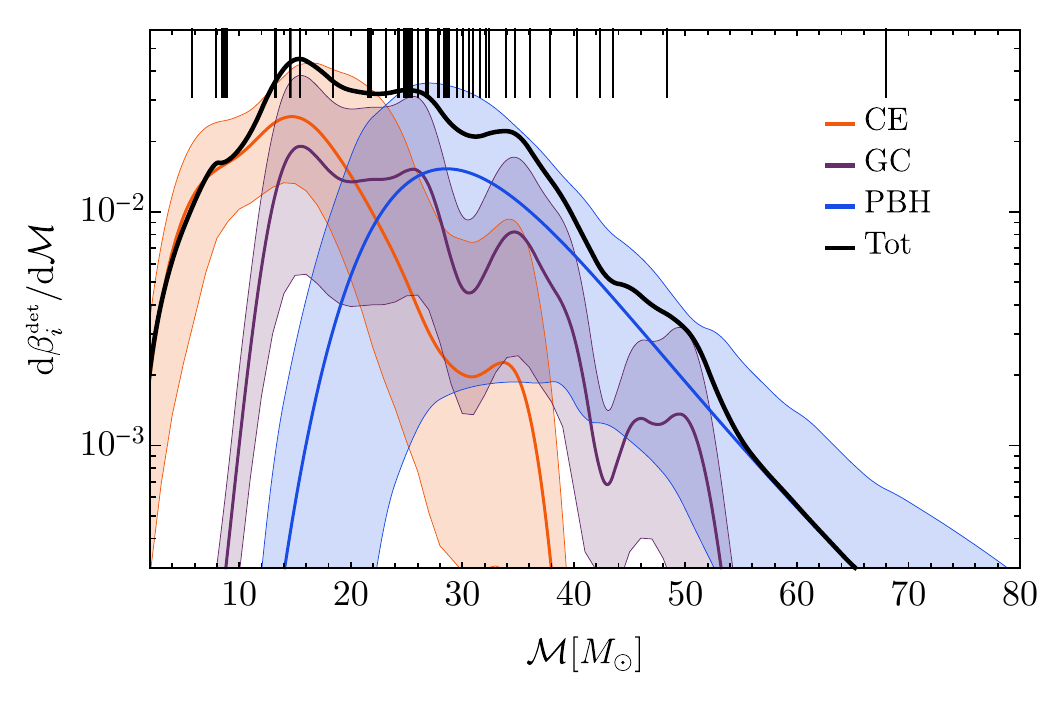}
\includegraphics[width=0.475 \linewidth]{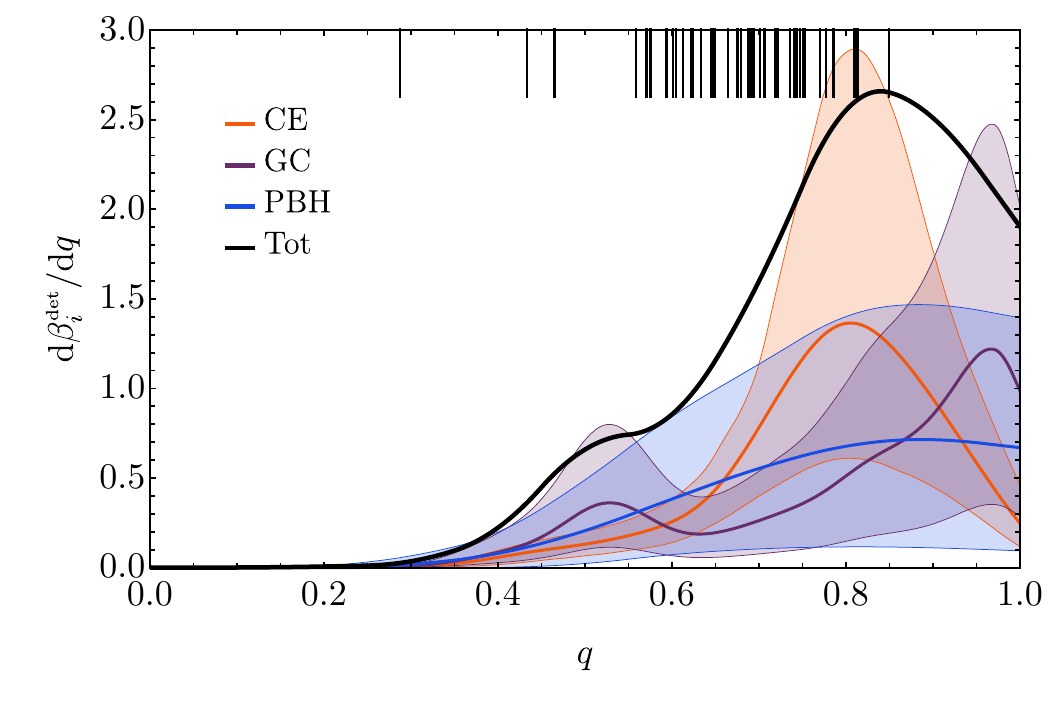}
\captionsetup{justification=centering, singlelinecheck=false}
	\caption{CE+GC+PBH mixed scenario including GW190521.}
	\label{fig:PDFseparate_7a}
	\end{subfigure}
	\begin{subfigure}[b]{1\textwidth}
          \centering
\includegraphics[width=0.475\linewidth]{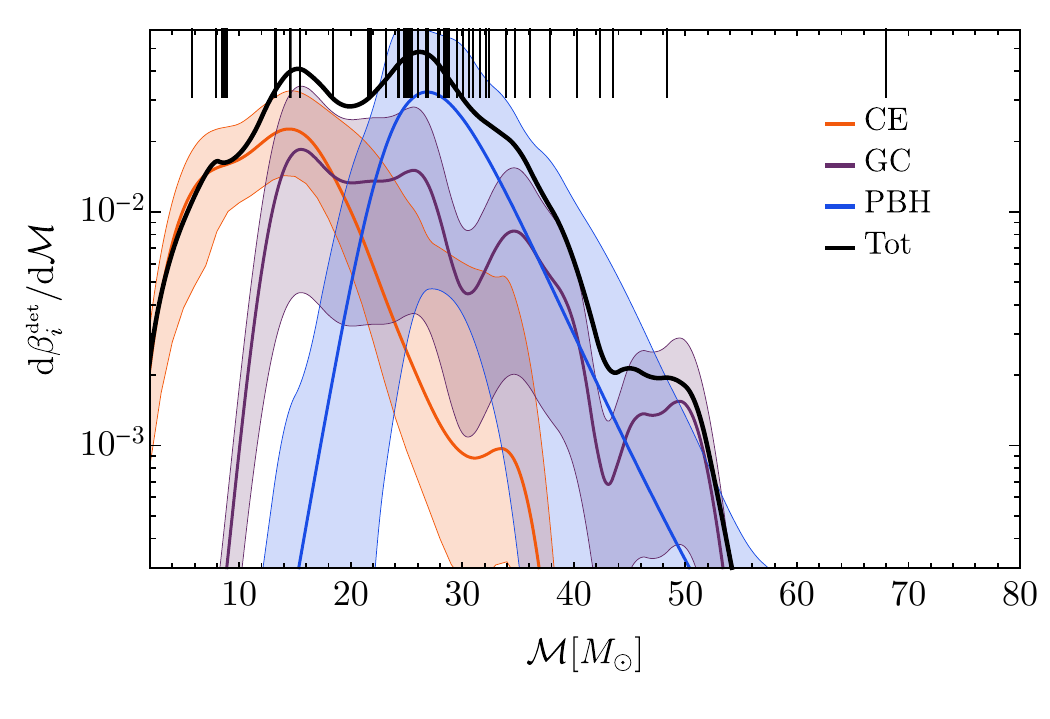}
\includegraphics[width=0.475 \linewidth]{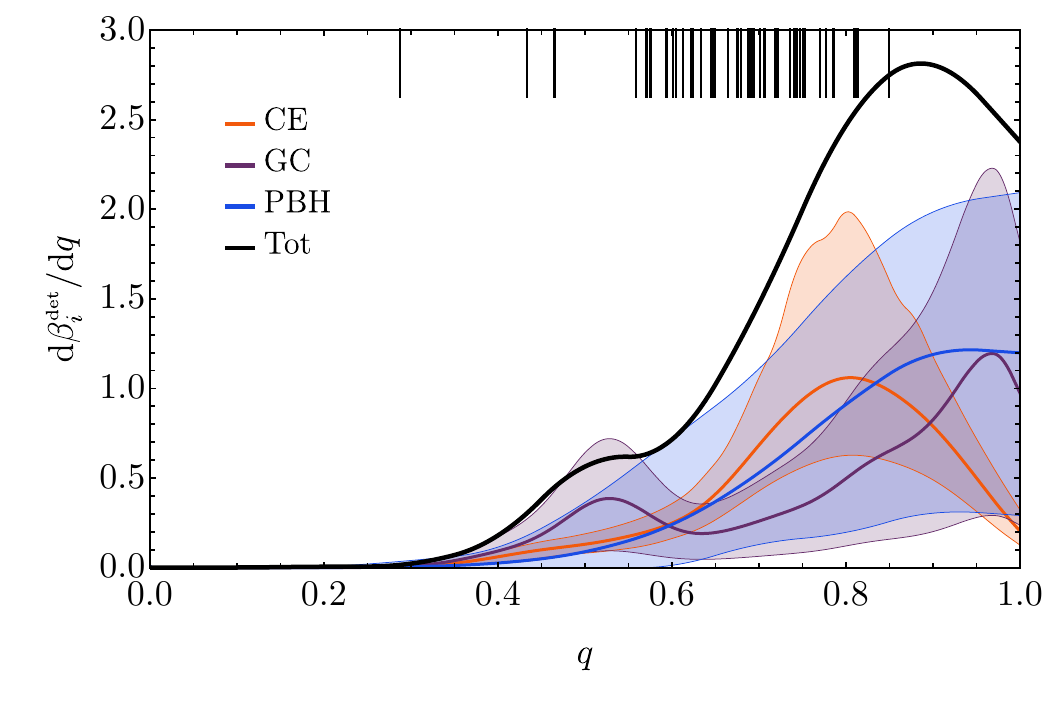}
\captionsetup{justification=centering, singlelinecheck=false}
\caption{ CE+GC+PBH mixed scenario without including GW190521.}
	\label{fig:PDFseparate_7b}
\end{subfigure}
\caption{
Same as Fig.~\ref{fig:PDFseparate_6} for the 2+1 mixed scenario with and without GW190521.}
\label{fig:PDFseparate_7}
\end{figure*}

In the two cases when the SMT channel is included, the PBH population parameters are less constrained. If we consider GW190521 as part of the data set, the small fraction of observable events ascribed to the PBH sector implies that constraints on the PBH sub-population come almost exclusively from GW190521. As a consequence, the population
is in general shifted to heavier masses, but with relatively larger uncertainties on the mass function parameters. Also, $f_\PBH$ peaks at slightly smaller values compared to the other scenarios.
If instead we discard GW190521, the posterior of  $f_\PBH$ has a plateau, reaching values compatible with zero. 
This confirms the importance of mass-gap events for constraining the PBH contribution to the observed GW events.
In all cases, the inferred observable fraction $\beta_\PBH^\text{\tiny det}$ forces the PBH abundance to be below $f_\PBH \lesssim 10^{-3}$, confirming that PBHs in the mass range currently observed by the LIGO and Virgo experiments can only be a small fraction of the dark matter.

In Fig.~\ref{fig:PDFseparate_5} we complement the chirp mass distributions shown in Fig.~\ref{fig:PDFseparate} by displaying the mass ratio distributions for the same 3+1 scenarios, including GW190521.
The chirp mass and mass ratio distributions that result from neglecting GW190521 are instead shown in Fig.~\ref{fig:PDFseparate_6}. In particular, by comparing Figs.~\ref{fig:PDFseparate_6a} and \ref{fig:PDFseparate_6b} with Fig.~\ref{fig:PDFseparate} we see that the inclusion of GW190521 shifts the PBH chirp mass distribution to higher values. The PBH mass distribution never strongly overlaps with the CE channel, so there is no degeneracy between the two populations, while the PBH channel is somewhat correlated with the dynamical channels and with the SMT channels (when included). 
In the CE+GC+SMT+PBH scenario without the mass gap event, the fraction $\beta_\PBH^\text{\tiny det}$ is compatible with zero.
This also implies that the PBH contribution shown in Fig.~\ref{fig:PDFseparate_6b} is not bounded from below.
In Fig.~\ref{fig:PDFseparate_7} we show the analogous distributions in the 2+1 scenario, with (top panel) and without (bottom panel) GW190521.
Note that distributions of $q$ are overall very similar to each other, peaking close to $q\simeq1$.
Only the NSC channel is characterized by a bimodal distribution (see Ref.~\cite{Zevin:2020gbd}).

\bibliography{main}

\end{document}